\documentclass{ifacconf}

\usepackage{graphicx}      %
\usepackage{natbib}        %

\usepackage{amsmath}       %
\usepackage{amssymb}       %
\usepackage{amsfonts}      %
\usepackage{siunitx}       %
\usepackage[mathscr]{euscript}
\usepackage{xcolor}        %
\usepackage{layouts}
\usepackage{microtype}

\newif\ifhighlightchanges

\ifhighlightchanges
\newcommand{\highlightred}[1]{\textcolor{red}{#1}}    %
\else

\newcommand{\highlightred}[1]{#1}
\fi

\newcommand{\mm}[1]{\boldsymbol{#1}} %

\newif\ifarxivversion
\arxivversiontrue %

\graphicspath{{./images/}} %

\begin{document}
\ifarxivversion
\thispagestyle{empty}
\vspace*{2cm}
\noindent{\LARGE IFAC Copyright Notice}\\[1em]
\fboxrule=0.4pt \fboxsep=8pt
\noindent\fbox{\begin{minipage}{\dimexpr\linewidth-2\fboxsep-2\fboxrule\relax}
		\textcopyright~2026 the authors. This work has been accepted to IFAC 
		for publication under a Creative Commons Licence CC-BY-NC-ND.\\[0.8em]
		
		Accepted to be published in: IFAC-PapersOnLine, Proceedings of the 
		24th IFAC World Congress, Busan, Republic of Korea, August 2026.\\[0.8em]
		
		DOI: to be added upon publication.
\end{minipage}}
\newpage
\fi
\begin{frontmatter}

\title{Neural Network-Based Virtual Wheel-Speed Sensor for Enhanced Low-Velocity State Estimation\thanksref{footnoteinfo}}

\thanks[footnoteinfo]{This research project was supported by IAV GmbH and the Institute of Automotive Engineering (IAE), TU Braunschweig}

\author[IMES]{Hendrik Sch\"afke} 
\author[IMES]{Daniel O. M. Weber}
\author[IAV]{Askar Vagapov}
\author[IAV]{Christoph Schweers}
\author[IMES]{Thomas Seel} 
\author[IMES]{Simon F. G. Ehlers}

\address[IMES]{Leibniz University Hannover, Institute of Mechatronic Systems, 30823 Garbsen, Germany (e-mail: schaefke@imes.uni-hannover.de).}
\address[IAV]{IAV GmbH, 38518 Gifhorn, Germany}

\begin{abstract} %
Accurate wheel speed information is crucial for vehicle control and state estimation.
Conventional sensors suffer from quantization and latency, especially at low velocities, while motor-speed signals in electric vehicles are distorted by drivetrain torsion.
This work presents a neural-network-based virtual wheel-speed sensor that fuses wheel-speed and motor-speed signals to reduce errors from both sources.
Validated on real-world Volkswagen ID.7 data, the real-time capable model achieves an error reduction of up to~85\% compared to the production sensor and~47\% compared to an optimized zero-phase filter, providing a smooth signal for driver-assistance functions.
The results demonstrate robust generalization across diverse real-world maneuvers within the vehicle platform.
\end{abstract}


\end{frontmatter}
\section{Introduction}
\begin{figure}[t]
	\includegraphics[width=0.956\columnwidth, trim=0mm 0.4mm 0mm 0.3mm, clip]{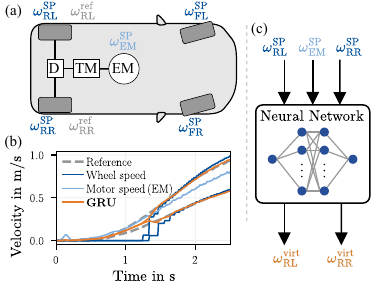}
	\caption{Overview of the proposed neural-network-based virtual wheel-speed sensor.
		(a) Vehicle schematic with drivetrain (D: differential, TM: transmission, EM: electric machine) and available rotational speed signals.
		(b) Low-speed behavior showing quantization and torsion-induced deviations, with left and right wheel speeds differing due to cornering.
		(c) Gated Recurrent Unit~(GRU) network architecture.
		}
	\label{fig:overview}
\end{figure}

Precise knowledge of the vehicle's state is essential for the functionality and safety of modern advanced driver-assistance systems~(ADAS).
Reviews emphasize that robust control performance critically depends on reliable state estimation~\citep{Review.Guo2018, Review.Jin2019}.
Since key variables like the sideslip angle cannot be measured directly with low-cost production sensors or are degraded by noise, state estimation has become standard practice in automotive engineering.
Research has therefore focused on estimating states like sideslip angle~\citep{Annunziata2024}, longitudinal velocity~\citep{Annunziata2022}, and tire-road friction~\citep{Schaefke2023} using model-based and data-driven methods.
Their performance, however, strongly depends on the quality of input signals such as the wheel speed.
At low speeds, conventional wheel-speed sensor signals are dominated by quantization and latency errors~\citep{Ragab2025}, limiting reliable estimation below \SI{3}{km/h}.
\highlightred{Crucial drivetrain functions like anti-jerk control are severely impaired by the poor quality or complete absence of wheel-speed signals at these low velocities~\citep{Review.Scamarcio2020}.}
Higher-resolution sensors could mitigate these issues, but their cost limits series production use.
In electric vehicles~(EVs), high-resolution motor-speed signals are readily available and offer a promising basis for improved wheel-speed estimation.
Unlike combustion-engine vehicles with a clutch-decoupled drivetrain, EVs feature a fixed connection between motor and wheels, so the motor-speed signal directly relates to wheel motion.
However, drivetrain torsion, backlash, and other nonlinearities distort this signal, especially during transient torque changes~\citep{Pu2023}.
Figure~\ref{fig:overview}a shows available series-production speed signals, while Fig.~\ref{fig:overview}b shows the characteristic low-speed behavior.
This challenge motivates the development of a virtual wheel-speed sensor that improves the quality of key input signals for existing estimation and control architectures, as illustrated in Fig.~\ref{fig:overview}c.

Vehicle state estimation approaches can generally be divided into model-based and data-driven methods. 
Model-based estimators, such as the Unscented Kalman filter~(UKF), rely on explicit physical vehicle models for sensor fusion~\citep{UKF.Julier1997}.
For example, \cite{Heidfeld2020} and \cite{Alshawi2024} presented UKF-based estimation of vehicle velocity and slip angle under varying road conditions.

\highlightred{While low-speed state estimation remains largely unaddressed in the literature, a relevant model-based approach published in a regional journal was introduced by \cite{Pu2023}, who proposed an interacting multiple-model~(IMM) estimator that fuses two velocity sources available in EVs: a wheel-speed–based model and a motor-speed–based model.}
The wheel-speed model provides robustness at moderate speeds but suffers from quantization and latency at low speeds, while the motor-speed model offers fast updates but is sensitive to drivetrain backlash and torsional oscillations. 
Although the IMM framework improves estimation continuity at low speeds, the drivetrain-induced disturbances affecting the motor-speed signal are not modeled and therefore remain uncompensated, leading to residual dynamic errors during transient torque changes.
Such disturbances are particularly problematic because robust longitudinal controllers, for example those used in adaptive cruise control, must suppress sensor-induced oscillations to maintain smooth acceleration tracking~\citep{Zhu2023}.
Consequently, the torsion- and backlash-induced fluctuations reported by~\cite{Pu2023} may propagate into low-speed control loops, for example during automated parking, leading to oscillatory or jerky vehicle motion.

Kalman filter–based estimators further assume additive white Gaussian noise, which fails to capture the structured, speed-dependent quantization and latency effects observed in wheel-speed measurements~\citep{Ragab2025}. 
\highlightred{While standard observers like the EKF or UKF are well-established for vehicle state estimation, applying them to this specific sensor fusion task would require a real-time capable analytical model capturing high-frequency drivetrain backlash and unmeasured external inputs such as tire forces.}
In practice, these nonlinear dynamics are difficult to model accurately using analytical formulations. 
\highlightred{To the best of the authors' knowledge, no such model-based approach addressing these combined challenges exists in the literature.}
This motivates the use of data-driven approaches that can directly learn nonlinear relationships and noise characteristics from sensor data.

While no existing work applies data-driven approaches to low-speed wheel-speed estimation or integrates EV drivetrain signals, neural networks have demonstrated strong performance in many related vehicle state estimation tasks. 
They have been used to estimate sideslip angle~\citep{Annunziata2024, Giuliacci2023,  Ziaukas2021, Srinivasan2020, Gonzalez2020, Ghosh2018}, to develop physically interpretable estimators for lateral velocity~\citep{Lio2023}, and to learn dynamic vehicle models capturing coupled longitudinal and lateral effects~\citep{Hermansdorfer2020}. 
Several studies have also addressed longitudinal velocity estimation~\citep{Annunziata2022, Bonfitto2021}. 
While the motor speed could in principle be used as an input for longitudinal velocity estimation, this would implicitly embed the dynamic tire radius, affected by wear, inflation pressure, and replacement, into the black-box model, which reduces robustness.
Similarly, although \cite{Onyekpe2021} used neural networks to correct wheel odometry for positioning, their method does not incorporate EV drivetrain signals.

In summary, low-speed state estimation remains insufficiently addressed.
This gap motivates the development of a virtual sensor that fuses wheel-speed, motor-speed, and torque signals to provide a more accurate estimate of the wheel speed to be used in ADAS.
The main contributions of this work are as follows:
\begin{itemize}
	\item Introduction of the first neural-network-based sensor-fusion approach that combines wheel-speed, motor-speed, and torque signals to form a virtual wheel-speed sensor mitigating drivetrain torsion, backlash, and wheel-speed discretization effects.
	\item Experimental validation on real-world data from a Volkswagen ID.7 across diverse driving scenarios.
	\item Demonstration of substantial error reductions compared with the production wheel-speed sensor and optimized low-pass filter baselines.
	\item Systematic investigation of model architectures, size, and computational cost, showing that compact models enable real-time deployment on automotive ECUs.
\end{itemize}
The remainder of this paper is structured as follows.
First, Section~\ref{sec:methods} introduces the neural network architectures used in this work.
The subsequent Section~\ref{sec:vehicle} provides an overview of the experimental vehicle, a Volkswagen ID.7, and the sensor setup.
Section~\ref{sec:results} then describes the experimental evaluation, where network sizes are systematically compared to assess model performance.
Finally, Section~\ref{sec:conclusion} concludes the paper with a summary of key findings and an outlook on future research directions.

\section{Methods}\label{sec:methods}
This section outlines the neural network architectures and optimization techniques employed in this work.

Recurrent neural networks are widely used for processing sequential data but often suffer from vanishing gradients when modeling long-term dependencies.
The Long Short-Term Memory~(LSTM) architecture~\citep{LSTM.Hochreiter1997} addresses this by introducing gate mechanisms that regulate the information flow over time.
The Gated Recurrent Unit~(GRU)~\citep{GRU.Cho2014} simplifies this structure by reducing the number of gates and merging cell state and hidden state, reducing computational complexity while preserving long-term dependencies.

The Temporal Convolutional Network~(TCN)~\citep{TCN.Lea2016} offers an alternative to recurrent architectures by employing dilated, causal convolutions and residual connections.
This structure enables modeling long-range temporal dependencies and allows for parallelized training.
Compared to recurrent models, TCNs can achieve improved convergence and scalability for longer input sequences.

\highlightred{In this work, LSTMs, GRUs, and TCNs are utilized to learn the mapping from the input vector $\mm{x}$ to the estimated wheel speeds $\hat{\mm{y}}$.}
\highlightred{Due to space constraints, we refer to \citep{GRU.Cho2014, TCN.Lea2016} for the detailed mathematical formulations of these standard architectures.}
For the recurrent approaches, each model consists of an input layer, a single recurrent layer, and a linear output layer that maps the hidden state to the predicted wheel-speed values. 
The TCN model uses a stack of dilated causal convolutional layers with exponentially increasing dilation factors $2^n$ in layer $n$, followed by a linear output layer. 
\highlightred{This setup follows standard non-autoregressive TCN designs \citep{Weber2021} to ensure that the estimate $\hat{\mm{y}}_t$ only depends on current and past inputs $\mm{x}_{\le t}$.}

All models were trained with Rectified Adam (RAdam)~\citep{RAdam.Liu2021}.
RAdam stabilizes early training by rectifying the variance of the adaptive learning rate, which mitigates the need for warm-up schedules and improves convergence robustness.
Hyperparameter optimization was performed using the Asynchronous Successive Halving Algorithm~(ASHA)~\citep{ASHA.Li2020}, enabling efficient exploration of many configurations by early-stopping unpromising trials and focusing on the best-performing candidates.

\section{Vehicle Platform and Dataset}\label{sec:vehicle}
The experimental data were collected with a Volkswagen ID.7 featuring an electric rear-wheel-drive powertrain.
All production signals, including motor speed, motor torque, brake torque, and wheel speeds, were recorded via the vehicle’s Controller Area Network~(CAN).
In the following, the series-production signals are referred to as SP.

\begin{figure}[t]
	\centering
	\begin{minipage}[b]{0.54\linewidth}
		\centering
		\includegraphics[width=\linewidth, trim=0 50 0 0, clip]{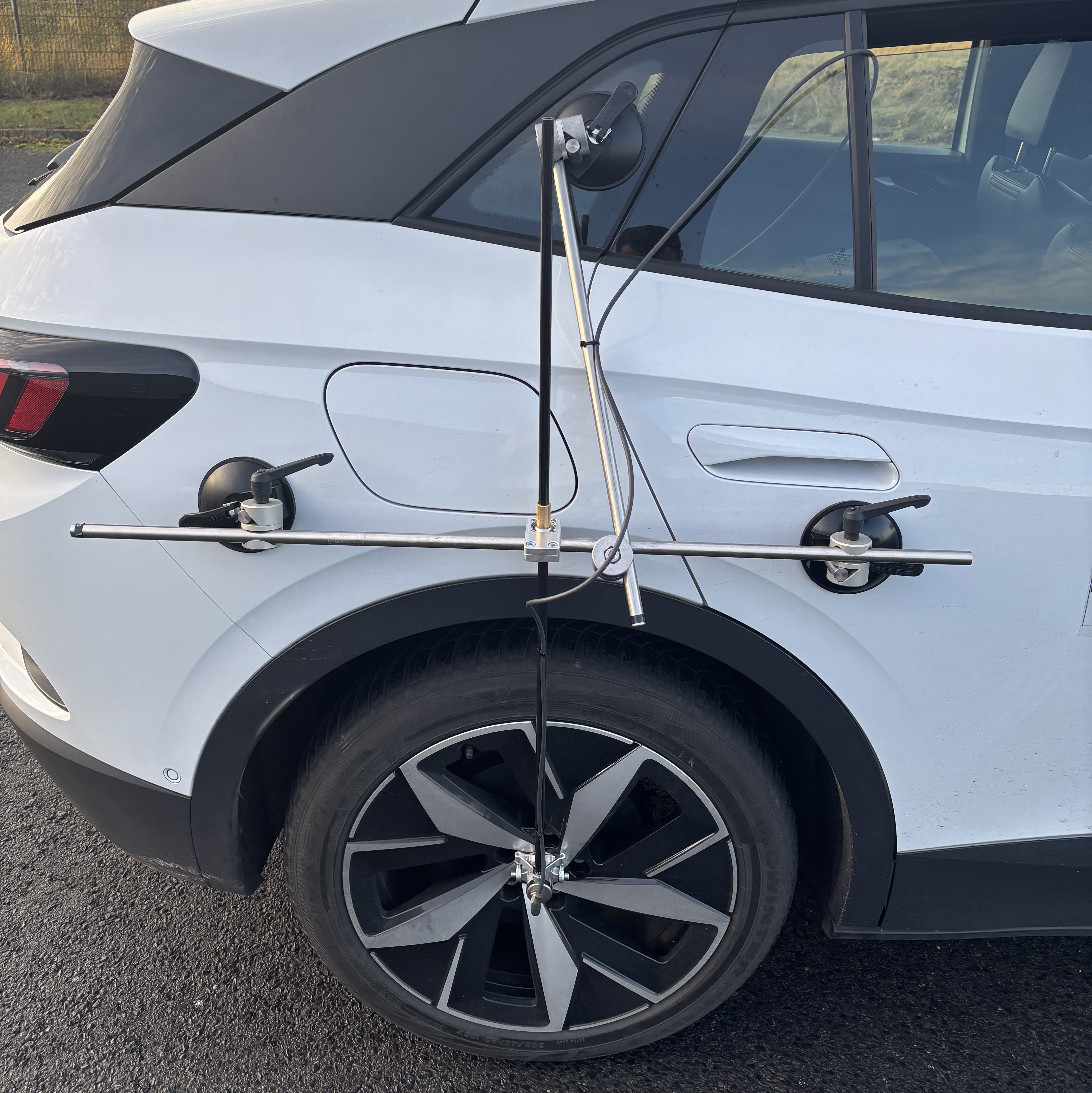}
		\vspace{1mm} %
		\centerline{(a)}
	\end{minipage}
	\hfill
	\begin{minipage}[b]{0.42\linewidth}
		\centering
		\includegraphics[width=\linewidth, trim=0 0 0 0, clip]{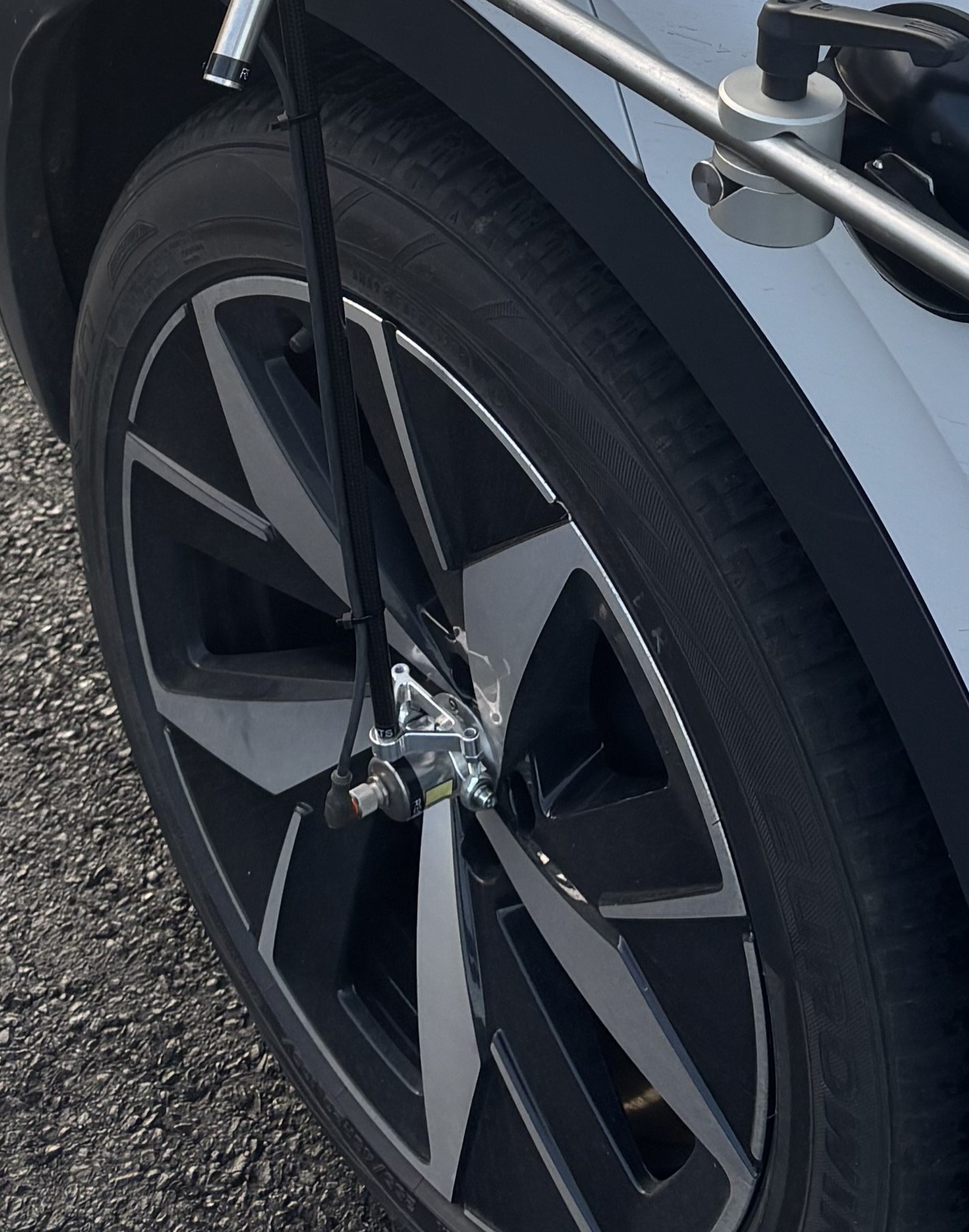}
		\vspace{1mm} %
		\centerline{(b)}
	\end{minipage}
	\caption{\highlightred{Experimental setup showing the external installation of the high-precision reference sensor on the rear axle to acquire ground-truth wheel-speed data: (a)~full view, (b)~zoomed-in view of the wheel center.}}
	\label{fig:raddrehzahlsensor}
\end{figure}

To obtain ground-truth data, the vehicle was equipped with two SIKO reference wheel-speed sensors mounted externally on the driven axle \highlightred{(see Fig.~\ref{fig:raddrehzahlsensor})}, each providing $4\,096$ ticks per revolution.
These sensors were also recorded via CAN.
For comparison, some series-production wheel-speed sensors with around $43$ ticks per revolution resolve wheel motion in steps of about \SI{4}{\centi\meter} to \SI{5}{\centi\meter} when converted into the corresponding vehicle displacement.
In contrast, the reference sensors used here provide a resolution on the order of \SI{0.5}{\milli\meter}.
This resolution gap justifies treating the reference signal as ground truth for this study.

For a better comparison, both the wheel-speed and motor-speed signals were converted to equivalent translational velocities at the tire-road interface.
This conversion was performed using the fixed transmission ratio of the drivetrain and a constant effective tire radius.

The dataset comprises one hour of recorded driving data covering a wide range of operating conditions and dynamic maneuvers, including standstill starts, emergency braking, and curb impacts, conducted on both asphalt and gravel surfaces.
Different torque levels and braking intensities were tested to capture a broad spectrum of drivetrain dynamics and transient behaviors.
All signals were time-synchronized and resampled to $\SI{50}{\hertz}$ using a zero-order hold to ensure consistent temporal alignment across modalities.

The data were divided into training, validation, and test subsets in a ratio of 70:20:10, separated by distinct driving maneuvers to avoid data leakage between sets. 
This ensures evaluation under unseen dynamic conditions, providing a fair assessment of the generalization capability. 
The resulting dataset forms a representative basis for investigating data-driven virtual wheel-speed estimation under realistic operating conditions.

\section{Experimental Setup and Results}\label{sec:results}
The neural network models were implemented in \texttt{PyTorch} using the \texttt{tsfast} library~\citep{tsfast}.
The loss function is the Mean Squared Error~(MSE) between the estimated \highlightred{wheel speeds $\hat{\mm{y}}$ and the targets $\mm{y}$} obtained from the high-precision sensors.
Various sensor configurations were evaluated during model development, including longitudinal and lateral acceleration, yaw rate, and front wheel speeds.
The final input vector~\(\mm{x}\) consists of the series-production rear wheel-speed signals~
\(\omega_{\mathrm{RL}}^{\mathrm{SP}}\) and \(\omega_{\mathrm{RR}}^{\mathrm{SP}}\), 
the motor speed~\(\omega_{\mathrm{EM}}^{\mathrm{SP}}\), and the drive and brake torques~\(M_{\mathrm{drive}}\) and \(M_{\mathrm{brake}}\).
The output vector~\(\mm{y}\) contains the corresponding reference wheel speeds~
\(\omega_{\mathrm{RL}}^{\mathrm{ref}}\) and \(\omega_{\mathrm{RR}}^{\mathrm{ref}}\),
so that the resulting input and output are given by
\begin{align}
	\mm{x} &= \bigl[\omega_{\mathrm{RL}}^{\mathrm{SP}},	\ \omega_{\mathrm{RR}}^{\mathrm{SP}},	\ \omega_{\mathrm{EM}}^{\mathrm{SP}},	\ M_{\mathrm{drive}},	\ M_{\mathrm{brake}} \bigr]^{\mathrm{T}}, \\[2mm]
	\mm{y} &= \bigl[\omega_{\mathrm{RL}}^{\mathrm{ref}},\ \omega_{\mathrm{RR}}^{\mathrm{ref}}\bigr]^{\mathrm{T}}.
\end{align}

To assess the performance of the proposed virtual sensor, two baseline methods were considered:
(1) a causal low-pass filter (LPF) applied to the motor speed signal, and
(2) an acausal (zero-phase) LPF implemented with \texttt{filtfilt}.
Both LPFs aim to suppress the high-frequency disturbances induced by drivetrain torsion and backlash.
In each case, the LPF corresponds to a Butterworth filter whose order and cutoff frequency were optimized on the validation dataset using Particle Swarm Optimization.
For the acausal LPF, an additional temporal shift between $-10$ and $+10$ time steps was optimized to allow proper alignment with future samples.
In addition, an optimistic reference was obtained by re-optimizing the acausal LPF directly on the test dataset, again including both filter parameters and the temporal shift.
This variant is denoted as LPF (acausal) (test-data).
Since both acausal LPF variants rely on future data, they cannot be deployed in real time and serve only as upper bounds for LPF performance.

\begin{figure}[t]
	\centering
	\includegraphics[width=\columnwidth]{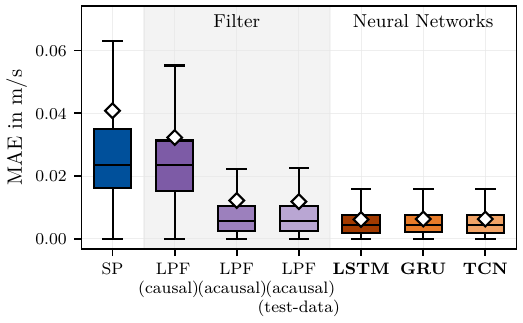}
	\caption{Comparison of the mean absolute error (MAE) of the wheel-speed signal relative to a reference sensor on an independent test dataset.
		The SP wheel-speed signal is shown on the left, the filter baselines (LPF) are highlighted by the shaded background, and the neural network approaches are shown on the right.
		The mean value of each distribution is indicated by a white diamond.
		Learning-based models outperform the production sensor and LPF baselines.}
	\label{fig:plot_error_boxplot}
\end{figure}

\begin{figure*}[t]
	\centering
	\includegraphics[width=\textwidth, trim=0mm 1.4mm 0mm 0.8mm, clip]{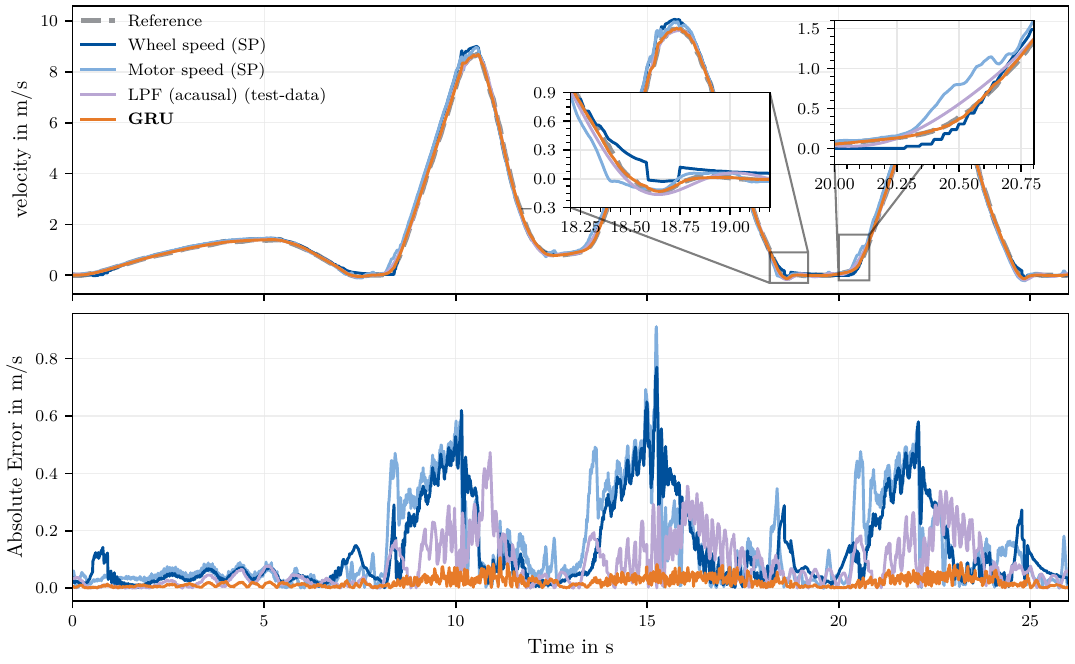}
	\caption{Comparison of the longitudinal velocity $v_x$ estimated by different methods (top) and the corresponding absolute error over time (bottom) of a test sequence.
		The SP wheel-speed signal shows clear inaccuracies at low speeds and during load transitions, while the GRU model achieves smoother and more accurate estimates.}
	\label{fig:plot_time_series_error}
\end{figure*}

All learning-based models were hyperparameter-optimized using ASHA on the validation set, during which the batch size, hidden size, and learning rate were automatically adapted.
The best-performing trial was subsequently evaluated on the independent test dataset. 
Each configuration was trained for at most 300~epochs, while ASHA terminated underperforming trials early. 
A cosine-annealing learning-rate schedule was used, and 100~hyperparameter combinations were sampled per network architecture \highlightred{with a fixed window size of 200 time steps.}
Performance was quantified using the Mean Absolute Error~(MAE) between the estimated and reference wheel speeds. 
Since the LPF baselines produce a single output signal, their performance was evaluated against the mean of the two high-precision reference wheel speeds on the driven axle. 
The two reference signals differ during cornering because the inner and outer wheels follow different path lengths.
A comparative overview of the MAE across all methods is provided in Fig.~\ref{fig:plot_error_boxplot}.
The boxplot illustrates the error distributions for the raw SP wheel-speed signal, the LPF motor speed baselines, and the neural network models (LSTM, GRU, and TCN).

The causal LPF, which represents a configuration that could in principle be deployed in a vehicle, reduced the MAE from \SI{0.0408}{m/s} for the raw SP wheel-speed signal to \SI{0.0323}{m/s}.
The acausal zero-phase LPF further reduced the MAE to \SI{0.0122}{m/s}, and additional tuning on the test dataset reduced it to \SI{0.0119}{m/s}.
The neural network models achieved the best performance, reaching an MAE of approximately \SI{0.006}{m/s}, corresponding to an error reduction of~84.5\% over the SP wheel-speed signal,~80.4\% over the causal LPF, and~46.7\% compared to the best acausal LPF.
All neural network architectures exhibited similar accuracy, with marginal MAE differences.
Among them, the GRU model provided the lowest computational cost and was selected for further analysis.

Figure~\ref{fig:plot_time_series_error} shows the time-domain comparison for a representative maneuver involving dynamic acceleration and deceleration with load changes and braking phases.
For clarity, only the GRU model and the acausal baseline (zero-phase LPF) are shown.
The rear-left~(RL) and rear-right~(RR) wheel speeds were averaged for both the reference and GRU outputs to improve readability.
The maneuver was recorded during a straight-line drive on dry asphalt.

Across all methods, the largest estimation errors occur primarily during load transitions, when drivetrain torsion and backlash induce oscillations between motor and wheel speeds.
These transient deviations, visible, for example, around seconds 10, 15, and 23, can only be partially compensated by the acausal LPF baseline.
Two zoomed-in sections in the upper subplot further illustrate the model behavior during braking and acceleration phases.
In both cases, the wheel-speed signal (dark blue) exhibits delayed and quantized behavior, while the motor speed (light blue) is distorted by torsional oscillations and backlash.
The GRU model provides accurate and dynamically consistent estimates in these regions, whereas the filtered baseline still produces residual offsets, such as nonzero velocity values during standstill.
Overall, the proposed neural virtual sensor employs sensor fusion to model and mitigate the nonlinear coupling between motor and wheel dynamics, resulting in accurate and physically consistent estimates even under rapidly changing load conditions.

\begin{figure}[]
	\centering
	\includegraphics[width=\columnwidth, trim=0mm 1.5mm 0mm 0.5mm, clip]{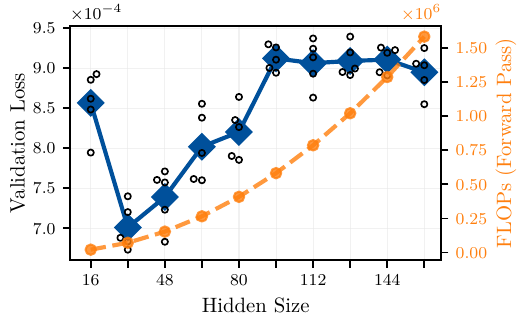}
	\caption{Validation loss (blue) and computational cost (orange) of single-layer GRU with varying hidden sizes. The black circles indicate individual trials, while the blue diamonds show the mean across the trials.}%
	\label{fig:plot_hidden_size}
\end{figure}

In addition to numerical accuracy, computational efficiency was evaluated with respect to potential real-time deployment.
A network complexity analysis was conducted by varying the GRU hidden size between 16 and 160 units in a single-layer configuration, training five independent models per configuration (see Fig.~\ref{fig:plot_hidden_size}).
The required floating-point operations (FLOPs) per forward pass are also reported, illustrating the trade-off between model performance and computational cost.
The best-performing GRU model, with a hidden size of 32, achieves the lowest validation loss while requiring only approximately 0.2~MFLOPs per inference.
Even assuming only one FLOP per clock cycle and core of an automotive-grade Infineon AURIX~TC3xx microcontroller at \SI{300}{\mega\hertz}, this corresponds to less than 0.1\% of the available throughput.
This suggests  that the proposed virtual wheel-speed sensor can be executed in real time on production electronic control units (ECUs) without the need for dedicated acceleration hardware.

The largest improvements occur under transient torque changes and at very low speeds, where drivetrain torsion and sensor quantization effects dominate.
Qualitatively, the neural model effectively suppresses torsional oscillations and produces a smooth, dynamically consistent wheel-speed signal suitable for real-time vehicle control applications.

\section{Conclusion and Future Work}\label{sec:conclusion}
This paper presented a neural-network-based virtual wheel-speed sensor designed to improve low-speed estimation accuracy in electric vehicles.
The proposed approach enhances performance by modeling the nonlinear relationship between motor and wheel-speed signals, including the effects of drivetrain torsion.
Three network architectures were systematically compared: Long Short-Term Memory~(LSTM), Gated Recurrent Unit~(GRU), and Temporal Convolutional Network~(TCN).
All learning-based models achieved high accuracy, with the GRU architecture providing the best trade-off between estimation performance and computational efficiency.
Due to its compact architecture, the GRU model can be executed efficiently on embedded vehicle hardware, enabling direct integration into production-grade estimation and control systems.

Compared with the production wheel-speed sensor, the virtual sensor achieves an error reduction of up to~85\%.
Even when compared to an optimized zero-phase low-pass filter, which is tuned on the test data and has access to future samples through the use of \texttt{filtfilt}, the neural approach still yields an additional improvement of about~47\%.
The resulting signal remains smooth and dynamically consistent across transient operating conditions, enabling reliable operation of safety-critical driver-assistance functions such as automated parking and traffic-jam assistance without requiring additional hardware sensors.

The results demonstrate that data-driven models can effectively capture nonlinear sensor errors and dynamic drivetrain effects that are difficult to describe explicitly with first-principle model-based approaches.
This highlights the strong potential of learning-based virtual sensors to significantly enhance the quality of signals from low-cost, production-grade sensors with minimal computational effort.
By improving the accuracy of existing hardware without requiring additional or more expensive components, the proposed approach enables cost-efficient deployment in series vehicles.
Overall, the presented concept is an important step toward robust vehicle state estimation across all operating conditions and has the potential to unlock the full capabilities of low-velocity control.

Future work will focus on integrating the proposed virtual sensing approach into existing state-estimation frameworks to improve the accuracy of downstream estimates such as longitudinal and lateral velocity.
The results demonstrate that the approach generalizes well across distinct test scenarios, suggesting strong robustness within a given vehicle platform.
Building on this, we will investigate the transferability of the trained models across different vehicle platforms and drivetrain configurations.
Another promising direction could be the use of attention mechanisms to increase model interpretability by highlighting which input features and temporal segments contribute most to the estimated wheel-speed signal.

\addtolength{\textheight}{-0.2cm}

\bibliography{literature}  %

\begin{thebibliography}{28}
\providecommand{\natexlab}[1]{#1}
\providecommand{\url}[1]{\texttt{#1}}
\providecommand{\urlprefix}{URL }
\expandafter\ifx\csname urlstyle\endcsname\relax
  \providecommand{\doi}[1]{doi:\discretionary{}{}{}#1}\else
  \providecommand{\doi}{doi:\discretionary{}{}{}\begingroup \urlstyle{rm}\Url}\fi

\bibitem[{Alshawi et~al.(2024)Alshawi, De~Pinto, Stano, van Aalst, Praet, Boulay, Ivone, Gruber, and Sorniotti}]{Alshawi2024}
Alshawi, A., De~Pinto, S., Stano, P., van Aalst, S., Praet, K., Boulay, E., Ivone, D., Gruber, P., and Sorniotti, A. (2024).
\newblock An adaptive unscented kalman filter for the estimation of the vehicle velocity components, slip angles, and slip ratios in extreme driving manoeuvres.
\newblock \emph{Sensors}, 24(2).
\newblock \doi{10.3390/s24020436}.

\bibitem[{Bonfitto and Feraco(2021)}]{Bonfitto2021}
Bonfitto, A. and Feraco, S. (2021).
\newblock A data-driven method for vehicle speed estimation.
\newblock \emph{Communications - Scientific letters of the University of Zilina}.
\newblock \doi{10.26552/com.C.2021.3.B165-B177}.

\bibitem[{Cho et~al.(2014)Cho, van Merri{\"e}nboer, Bahdanau, and Bengio}]{GRU.Cho2014}
Cho, K., van Merri{\"e}nboer, B., Bahdanau, D., and Bengio, Y. (2014).
\newblock On the properties of neural machine translation: Encoder--decoder approaches.
\newblock In \emph{Proc. SSST-8}, 103--111. ACL.
\newblock \doi{10.3115/v1/W14-4012}.

\bibitem[{Ghosh et~al.(2018)Ghosh, Tonoli, and Amati}]{Ghosh2018}
Ghosh, J., Tonoli, A., and Amati, N. (2018).
\newblock A deep learning based virtual sensor for vehicle sideslip angle estimation: Experimental results.
\newblock In \emph{SAE Technical Paper Series}.
\newblock \doi{10.4271/2018-01-1089}.

\bibitem[{Giuliacci et~al.(2023)Giuliacci, Ballesio, Fainello, Mair, and King}]{Giuliacci2023}
Giuliacci, T.A., Ballesio, S., Fainello, M., Mair, U., and King, J. (2023).
\newblock Recurrent neural network model for on-board estimation of the side-slip angle in a four-wheel drive and steering vehicle.
\newblock \emph{SAE International Journal of Passenger Vehicle Systems}, 17(15-17-01-0003), 37--48.

\bibitem[{Gonz{\'a}lez et~al.(2020)Gonz{\'a}lez, S{\'a}nchez, Garcia-Guzman, Boada, and Boada}]{Gonzalez2020}
Gonz{\'a}lez, L.P., S{\'a}nchez, S.S., Garcia-Guzman, J., Boada, M.J.L., and Boada, B.L. (2020).
\newblock Simultaneous estimation of vehicle roll and sideslip angles through a deep learning approach.
\newblock \emph{Sensors}, 20(13).
\newblock \doi{10.3390/s20133679}.

\bibitem[{Guo et~al.(2018)Guo, Cao, Chen, Lv, Wang, and Yang}]{Review.Guo2018}
Guo, H., Cao, D., Chen, H., Lv, C., Wang, H., and Yang, S. (2018).
\newblock Vehicle dynamic state estimation: state of the art schemes and perspectives.
\newblock \emph{IEEE/CAA Journal of Automatica Sinica}, 5(2), 418--431.
\newblock \doi{10.1109/JAS.2017.7510811}.

\bibitem[{Heidfeld et~al.(2020)Heidfeld, Schünemann, and Kasper}]{Heidfeld2020}
Heidfeld, H., Schünemann, M., and Kasper, R. (2020).
\newblock Ukf-based state and tire slip estimation for a 4wd electric vehicle.
\newblock \emph{Vehicle System Dynamics}, 58(10), 1479--1496.
\newblock \doi{10.1080/00423114.2019.1648836}.

\bibitem[{Hermansdorfer et~al.(2020)Hermansdorfer, Trauth, Betz, and Lienkamp}]{Hermansdorfer2020}
Hermansdorfer, L., Trauth, R., Betz, J., and Lienkamp, M. (2020).
\newblock End-to-end neural network for vehicle dynamics modeling.
\newblock In \emph{2020 6th IEEE Congress on Information Science and Technology (CiSt)}, 407--412.
\newblock \doi{10.1109/CiSt49399.2021.9357196}.

\bibitem[{Hochreiter and Schmidhuber(1997)}]{LSTM.Hochreiter1997}
Hochreiter, S. and Schmidhuber, J. (1997).
\newblock Long short-term memory.
\newblock \emph{Neural computation}, 9(8), 1735--1780.
\newblock \doi{10.1162/neco.1997.9.8.1735}.

\bibitem[{Jin et~al.(2019)Jin, Yin, and Chen}]{Review.Jin2019}
Jin, X., Yin, G., and Chen, N. (2019).
\newblock Advanced estimation techniques for vehicle system dynamic state: A survey.
\newblock \emph{Sensors}, 19(19), 4289.
\newblock \doi{10.3390/s19194289}.

\bibitem[{Julier and Uhlmann(1997)}]{UKF.Julier1997}
Julier, S.J. and Uhlmann, J.K. (1997).
\newblock New extension of the kalman filter to nonlinear systems.
\newblock In \emph{SPIE Proceeding}, volume 3068, 182--193.
\newblock \doi{10.1117/12.280797}.

\bibitem[{Lea et~al.(2016)Lea, Flynn, Vidal, Reiter, and Hager}]{TCN.Lea2016}
Lea, C., Flynn, M.D., Vidal, R., Reiter, A., and Hager, G.D. (2016).
\newblock Temporal convolutional networks for action segmentation and detection.

\bibitem[{Li et~al.(2020)Li, Jamieson, Rostamizadeh, Gonina, Ben-tzur, Hardt, Recht, and Talwalkar}]{ASHA.Li2020}
Li, L., Jamieson, K., Rostamizadeh, A., Gonina, E., Ben-tzur, J., Hardt, M., Recht, B., and Talwalkar, A. (2020).
\newblock A system for massively parallel hyperparameter tuning.
\newblock In \emph{Third Conference on Systems and Machine Learning}.

\bibitem[{Lio et~al.(2023)Lio, Piccinini, and Biral}]{Lio2023}
Lio, M.D., Piccinini, M., and Biral, F. (2023).
\newblock Robust and sample-efficient estimation of vehicle lateral velocity using neural networks with explainable structure informed by kinematic principles.
\newblock \emph{IEEE Transactions on Intelligent Transportation Systems}, 24(12), 13670--13684.
\newblock \doi{10.1109/TITS.2023.3303776}.

\bibitem[{Liu et~al.(2019)Liu, Jiang, He, Chen, Liu, Gao, and Han}]{RAdam.Liu2021}
Liu, L., Jiang, H., He, P., Chen, W., Liu, X., Gao, J., and Han, J. (2019).
\newblock On the variance of the adaptive learning rate and beyond.
\newblock \urlprefix\url{http://arxiv.org/abs/1908.03265}.

\bibitem[{Napolitano~Dell’Annunziata et~al.(2022)Napolitano~Dell’Annunziata, Arricale, Farroni, Genovese, Pasquino, and Tranquillo}]{Annunziata2022}
Napolitano~Dell’Annunziata, G., Arricale, V.M., Farroni, F., Genovese, A., Pasquino, N., and Tranquillo, G. (2022).
\newblock Estimation of vehicle longitudinal velocity with artificial neural network.
\newblock \emph{Sensors}, 22(23).
\newblock \doi{10.3390/s22239516}.

\bibitem[{Napolitano~Dell’Annunziata et~al.(2024)Napolitano~Dell’Annunziata, Ruffini, Stefanelli, Adiletta, Fichera, and Timpone}]{Annunziata2024}
Napolitano~Dell’Annunziata, G., Ruffini, M., Stefanelli, R., Adiletta, G., Fichera, G., and Timpone, F. (2024).
\newblock Four-wheeled vehicle sideslip angle estimation: A machine learning-based technique for real-time virtual sensor development.
\newblock \emph{Applied Sciences}, 14(3).
\newblock \doi{10.3390/app14031036}.

\bibitem[{Onyekpe et~al.(2021)Onyekpe, Palade, Herath, Kanarachos, and Fitzpatrick}]{Onyekpe2021}
Onyekpe, U., Palade, V., Herath, A., Kanarachos, S., and Fitzpatrick, M.E. (2021).
\newblock Whonet: Wheel odometry neural network for vehicular localisation in gnss-deprived environments.
\newblock \emph{Engineering Applications of Artificial Intelligence}, 105, 104421.
\newblock \doi{https://doi.org/10.1016/j.engappai.2021.104421}.

\bibitem[{Pu et~al.(2023)Pu, Tang, Shangguan, Wang, and Jiang}]{Pu2023}
Pu, Z., Tang, L., Shangguan, W., Wang, W., and Jiang, K. (2023).
\newblock Research on the estimation of vehicle speed under low-speed conditions based on multi-sensor information.
\newblock \emph{Automotive Engineering}, 45(7), 1235--1243.

\bibitem[{Ragab et~al.(2025)Ragab, Givigi, and Noureldin}]{Ragab2025}
Ragab, H., Givigi, S., and Noureldin, A. (2025).
\newblock Automotive speed estimation: Sensor types and error characteristics from obd-ii to adas.
\newblock In \emph{2025 IEEE/ION Position, Location and Navigation Symposium (PLANS)}, 124--130.
\newblock \doi{10.1109/PLANS61210.2025.11028310}.

\bibitem[{Scamarcio et~al.(2020)Scamarcio, Gruber, {De Pinto}, and Sorniotti}]{Review.Scamarcio2020}
Scamarcio, A., Gruber, P., {De Pinto}, S., and Sorniotti, A. (2020).
\newblock Anti-jerk controllers for automotive applications: A review.
\newblock \emph{Annual Reviews in Control}, 50, 174--189.
\newblock \doi{https://doi.org/10.1016/j.arcontrol.2020.04.013}.

\bibitem[{Sch{\"a}fke et~al.(2023)Sch{\"a}fke, Lampe, and Kortmann}]{Schaefke2023}
Sch{\"a}fke, H., Lampe, N., and Kortmann, K.P. (2023).
\newblock Transformer neural networks for maximum friction coefficient estimation of tire-road contact using onboard vehicle sensors.
\newblock In \emph{2023 62nd IEEE Conference on Decision and Control (CDC)}, 5331--5338.
\newblock \doi{10.1109/CDC49753.2023.10384175}.

\bibitem[{Srinivasan et~al.(2020)Srinivasan, Sa, Zyner, Reijgwart, Valls, and Siegwart}]{Srinivasan2020}
Srinivasan, S., Sa, I., Zyner, A., Reijgwart, V., Valls, M.I., and Siegwart, R. (2020).
\newblock End-to-end velocity estimation for autonomous racing.
\newblock \emph{IEEE Robotics and Automation Letters}, 5(4), 6869--6875.
\newblock \doi{10.1109/LRA.2020.3016929}.

\bibitem[{Weber and Gühmann(2021)}]{Weber2021}
Weber, D. and Gühmann, C. (2021).
\newblock Non-autoregressive vs autoregressive neural networks for system identification.
\newblock \emph{IFAC-PapersOnLine}, 54(20), 692--698.
\newblock \doi{10.1016/j.ifacol.2021.11.252}.
\newblock Modeling, Estimation and Control Conference MECC 2021.

\bibitem[{Weber(2024)}]{tsfast}
Weber, D.O. (2024).
\newblock tsfast - a deep learning library for time series analysis and system identification.
\newblock Github.
\newblock \urlprefix\url{https://github.com/daniel-om-weber/tsfast}.

\bibitem[{Zhu et~al.(2023)Zhu, Bei, Li, Liu, Tang, Zhu, and Gao}]{Zhu2023}
Zhu, Z., Bei, S., Li, B., Liu, G., Tang, H., Zhu, Y., and Gao, C. (2023).
\newblock Research on robust control of intelligent vehicle adaptive cruise.
\newblock \emph{World Electric Vehicle Journal}, 14(10).
\newblock \doi{10.3390/wevj14100268}.

\bibitem[{Ziaukas et~al.(2021)Ziaukas, Busch, and Wielitzka}]{Ziaukas2021}
Ziaukas, Z., Busch, A., and Wielitzka, M. (2021).
\newblock Estimation of vehicle side-slip angle at varying road friction coefficients using a recurrent artificial neural network.
\newblock In \emph{2021 IEEE Conference on Control Technology and Applications (CCTA)}, 986--991.
\newblock \doi{10.1109/CCTA48906.2021.9658710}.

\end{thebibliography}

\end{document}